\documentclass{article}
\usepackage{dcase2020,amsmath,graphicx,url,times,booktabs, tabularx, array}
\usepackage{color}
\usepackage{comment}
\usepackage{multirow}
\usepackage{setspace}

\newlength\savedwidth

\title{RWCP-SSD-Onomatopoeia: onomatopoeic word dataset for environmental sound synthesis}

\name{Yuki Okamoto$^{1}$,
       Keisuke Imoto$^{2, 1}$,
       Shinnosuke Takamichi$^{3}$, 
       }
 \secondlinename{	  
       Ryosuke Yamanishi$^{1, 4}$, 
       Takahiro Fukumori$^{1}$, 
       Yoichi Yamashita$^{1}$
       }
 \address{$^1$ Ritsumeikan University, Japan, $^2$ Doshisha University, Japan, \\$^3$ The University of Tokyo, Japan, $^4$ Kansai University, Japan\\
  }

\begin{document}

\ninept
\maketitle

\begin{sloppy}
\begin{abstract}
Environmental sound synthesis is a technique for generating a natural environmental sound.
Conventional work on environmental sound synthesis using sound event labels cannot finely control synthesized sounds, for example, the pitch and timbre.
We consider that onomatopoeic words can be used for environmental sound synthesis.
Onomatopoeic words are effective for explaining the feature of sounds. 
We believe that using onomatopoeic words will enable us to control the fine time--frequency structure of synthesized sounds.
However, there is no dataset available  for environmental sound synthesis using onomatopoeic words.
In this paper, we thus present RWCP-SSD-Onomatopoeia, a dataset consisting of 155,568 onomatopoeic words paired with audio samples for environmental sound synthesis.
We also collected self-reported confidence scores and others-reported acceptance scores of onomatopoeic words, to help us investigate the difficulty in the transcription and selection of a suitable word for environmental sound synthesis.
\end{abstract}
%
%
%
\vspace{-3pt}
\section{Introduction}
\label{sec:intro}
\vspace{-3pt}
Environmental sound synthesis is a new field of audio generation and is the task of generating a natural environmental sound. 
In many studies on environmental sound synthesis, a physical modeling approach has been taken \cite{Salamon_WASPAA2017_01, Schwarz_DAFx2011, Bernardes_SMCC2016_01}. 
In recent years, some methods of environmental sound synthesis based on statistical generative models such as the deep learning approach have been developed \cite{okamoto_arXiv_2019, Kong_ICASSP2019_01, Caracalla_ICASSP_2020, Liu_arXiv_2020}.
Environmental sound synthesis has the potential for many applications such as supporting movie and game production \cite{Kong_ICASSP2019_01, Lloyd_ACMI3DGG_01}, generation of content for virtual reality (VR) \cite{Wang_VR2017_01, Zhou_CVPR2018_01}, and data augmentation for sound event detection and scene classification \cite{Salamon_WASPAA2017_01, Gontier_ICASSP_2020}.
In methods of environmental sound synthesis, sound event synthesis (SES) using the sound event labels as the input of the system \cite{okamoto_arXiv_2019} has been proposed. 
However, using only sound event labels does not allow fine control of the time--frequency structure for synthesized sounds, such as the pitch and timbre. 

To control synthesized environmental sounds more finely, we can apply environmental sound synthesis using onomatopoeic words as the input of the system.
Since an onomatopoeic word is a character sequence that phonetically imitates a sound, the use of such words to control the time--frequency structure of the sound is reasonable.
For example, Lemaitre and Rocchesso \cite{Lemaitre_2018} and Sundaram and Narayanan \cite{Sundaram_2006} have shown that using onomatopoeic words is effective for expressing the features of audio samples.
Fig.~\ref{fig:SES_onomatopeia} shows an overview of environmental sound synthesis using onomatopoeic words.
To synthesize environmental sounds from onomatopoeic words, a relationship to model training between the environmental sound and onomatopoeic words must be obtained. 
Thus, a dataset of onomatopoeic words matched with environmental sounds is required.

In this paper, we present the freely available dataset RWCP-SSD-Onomatopoeia, for environmental sound synthesis using onomatopoeic words.
We collected onomatopoeic words of 105 kinds of sound (e.g., shaver sound, whistle sound) included in RWCP-SSD (Real World Computing Partnership-Sound Scene Database) \cite{Nakamura_LREC2000_01}.
By requesting crowdworkers to transcribe the sound they hear, we obtained onomatopoeic words by crowdsourcing.
In some cases, multiple onomatopoeic words are collected for one sound event. 
We also collected worker's self-reported confidence scores and others-reported acceptance scores for the onomatopoeic words.
These scores help us investigate the difficulty in the transcription and selection of a suitable word for environmental sound synthesis.
The rest of the paper is organized as follows.
In Sec.~\ref{sec:creation_dataset}, we describe the creation of RWCP-SSD-Onomatopoeia, i.e., the collection of onomatopoeic words. 
In Sec.~\ref{sec:analysis_onomatope}, we analyze collected onomatopoeic words.
Finally, we summarize and conclude this paper in Sec.~\ref{sec:conclusion}. 

\begin{figure}[t!]
\centering
\vspace{10pt}
\includegraphics[width=0.93\columnwidth]{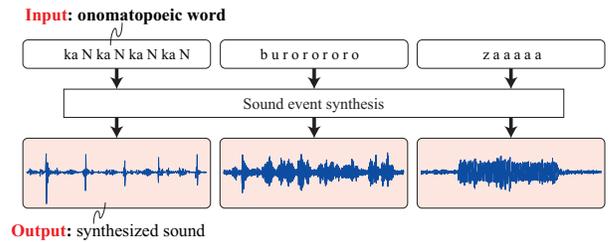}
\vspace{-5pt}
\caption{Overview of environmental sound synthesis using onomatopoeic words}
\label{fig:SES_onomatopeia}
\vspace{-5pt}
\end{figure}

\begin{table*}[t!]
\small
\caption{Examples of collected onomatopoeic words}
\vspace{2pt}
\label{table:exapmle_onomatopoeic}
\centering
\begin{tabular}{|l||l|l|l|l|} \hline
    \multirow{2}{*}{Sound event} & \multirow{2}{*}{Onomatopoeic word} & Self-reported & Others-reported & \multirow{2}{*}{Description of sound} \\ & & confidence score & acceptance score & \\  \hline 
             \ \\[-10pt]
             & p i i i i: & 4 & 4.9 & \\
    whistle1   & p i & 5 & 4.9 & Whistle-like sound with a constant high pitch\\
             & ts i: q & 1 & 2.8 & \\ \hline
             \ \\[-10pt]
            & b u: N & 4 & 4.5 & \\  
    shaver & {\bf j i:}  & {\bf 1} & {\bf 4.5} & Sound of operating an electric shaver\\
           & b u N b u N b u N &1 & 3.3 & \\\hline
             \ \\[-10pt]
         & hy u N q & 4 & 3.5 & \\
    file & {\bf m i: q }& {\bf 5 }&{\bf 1.9 }& Sound of rubbing a metal rod with a metal file\\
         & s a q: & 3 & 3.3 & \\ \hline
             \ \\[-10pt]
          & gy u r i gy u r i gy u r i & 4 & 4 & \\
    coffmill & g a r i g a r i g a r i & 5 & 3 & Sound by grinding beans in a coffee mill\\
         & b u b u b u b u b u & 3 & 3.3 & \\\hline
             \ \\[-10pt]
         & z u sh a a a a & 5 & 4.1 & \\
         tear  & b i y a b i y a & 3 & 2.9 & Sound by tearing paper\\
         & g i ry a g i ry a & 1 & 2.5 & \\ \hline
  \end{tabular}
  \vspace{8pt}
\end{table*}

\vspace{-3pt}
\section{CREATION OF RWCP-SSD-ONOMATOPOEIA}
\label{sec:creation_dataset}
\vspace{-3pt}
%
%
%
\vspace{-3pt}
\subsection{RWCP-SSD}
\label{sec:RWCP-SSD}
\vspace{-3pt}
We have collected onomatopoeic words for all nonspeech sounds in  RWCP-SSD.
RWCP-SSD contains 105 types of sound event, each of which includes about 100 audio samples (total of 9,722 audio samples).
Each audio sample is from 0.5 to 2.0 s in length. 
The sampling frequency is 48 kHz, and the quantization bit rate is 16 bits.
Included sound events are classified into the three categories in \cite{Nakamura_LREC2000_01} as follows:
\begin{itemize}
  \setlength{\itemsep}{0pt}
  \item{\bf Crash sounds}\\
  This class contains crash sounds of wood, metal, and plastic, such as the sound of a wooden board hit with a wooden stick.\\
  \item{\bf Sounds of human operation of objects easily associated with source materials}\\ 
  This class contains sounds of things being operated by humans, such as a whistle and telephone rings.\\[-6pt]
  \item{\bf Sounds of human operation of objects not easily associated with source materials}\\
  This class contains sounds of things being operated by humans, such as claps and sawing sounds.\\
\end{itemize} 
\vspace{-3pt}
%
%
\vspace{-15pt}
\subsection{Design of RWCP-SSD-Onomatopoeia}
\vspace{0pt}
%
The RWCP-SSD-Onomatopoeia dataset consists of the following contents.
\begin{itemize}
  \setlength{\itemsep}{0pt}
  \item{\bf Onomatopoeic words for each audio sample}\\
  We collected a total of 155,568 onomatopoeic words (9,722 audio samples $\times$ 5 or more people per audio sample; each crowdworker gave three different kinds of onomatopoeic words). 
  Each onomatopoeic word was collected from Japanese speakers in {\it katakana}, which is a Japanese syllabary, and was converted to the phoneme representation, which follows the conversion rule of Speech Segmentation Toolkit in the speech recognition engine Julius \cite{Julias}. \\
  \item{\bf Self-reported confidence score}\\
  We asked crowdworkers to score a confidence level for words they themselves transcribed.
  The self-reported confidence score enables us to evaluate the appropriateness of onomatopoeic words on the basis of the judgement of the person giving the onomatopoeic words.
  We describe the details of the self-reported confidence score in Sec.~\ref{sec:onomatope_collection}.\\
  \item{\bf Others-reported acceptance score}\\
  We asked crowdworkers to score an acceptance level for words transcribed by others.
  The others-reported acceptance score enables us to evaluate the appropriateness of onomatopoeic words on the judgement of others.
  We describe the details of the others-reported acceptance score in Sec.~\ref{sec:onomatope_collection}.\\
  \item{\bf Worker ID}\\
  The dataset includes anonymized IDs of crowdworkers who gave onomatopoeic words, confidence scores, and acceptance scores. \\[-8pt]
\end{itemize}
\begin{figure}[t!]
\includegraphics[width=1.0\columnwidth]{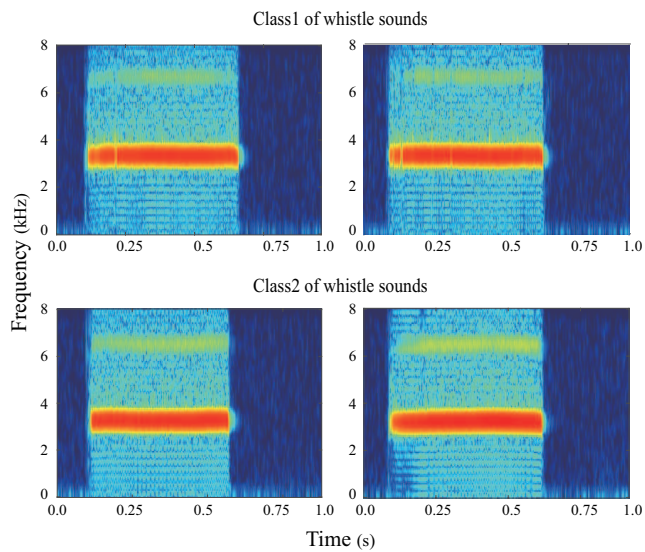}
\vspace{-21pt}
\caption{Spectrograms of whistle sounds}
\label{fig:whistle_spectrogram}
\vspace{-18pt}
\end{figure}
This dataset is freely available online\footnote{https://www.ksuke.net/dataset}. 
Note that RWCP-SSD-Onomatopoeia does not contain sound files, which can be obtained from Speech Resources Consortium (NII-SRC)\footnote{http://research.nii.ac.jp/src/en/index.html}.
These two resources have the same directory structure, and we can easily merge these resources.
%
%
\vspace{-3pt}
\subsection{Clustering of Audio Samples in Sound Events}
\label{sec:Clustering}
\vspace{-3pt}
%
In RWCP-SSD, each sound event contains about 100 audio samples, some of which are similar. Fig.~\ref{fig:whistle_spectrogram} shows spectrograms of whistle sounds; class1 and class2 show a pair of similar sounds.
We believe that the same onomatopoetic words may be given to sounds having similar acoustic features.
Therefore, we classified audio samples into classes of similar sounds for each sound event. 
From each class, we selected one audio sample and assigned onomatopoeic words. 

To classify similar sounds for each sound event, we calculated the cross-correlation between waveforms in each sound event as 
\begin{equation}
\label{eq:cross-correlato}
R_{xy} = \frac{1}{\sqrt{R_{xx} R_{yy}}}R_{xy}
\end{equation}
where subscripts $x$ and $y$ indicate an audio sample in each sound event. 
$R_{xx}$ and $R_{yy}$ are autocorrelation coefficients. 
$R_{xy}$ indicates the cross-correlation coefficient for $x$ and $y$. 
If the cross-correlation coefficient is 0.5 or higher, the sounds are classified as being of the same class.
As a result of the classification using the cross-correlation, 9,722 audio samples were classified to 6,024 classes.

\begin{figure}[t!]
\includegraphics[width=1.0\columnwidth]{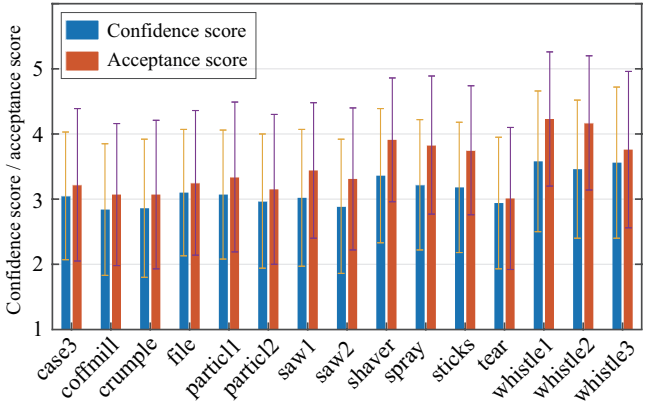}
\vspace{-20pt}
\caption{Average of self-reported confidence / others-reported acceptance scores for each sound event}
\label{fig:conf_acceptance}
\vspace{3pt}
\end{figure}

%
\vspace{-3pt}
\subsection{Onomatopoeic Word Collection}
\label{sec:onomatope_collection}
\vspace{-3pt}
We conducted the pre-experiment of collecting onomatopoeic words for the whistle sound.
In the pre-experiment, we collected different onomatopoeic words, such as ``p i i i i:'', ``p i'', and ``ts i: q'', for the same audio sample.
The pre-experimental results showed that multiple onomatopoeic words are collected for the same audio sample. 
We also collected self-reported confidence scores and others-reported acceptance scores, which help us investigate the difficulty in the transcription and selection of a suitable word for environmental sound synthesis.
The others-reported acceptance scores were collected for onomatopoeic words with a confidence level of 4 or high.

We collected onomatopoeic words, self-reported confidence scores, and others-reported acceptance scores for 6,024 audio samples from Japanese speakers.
From the results, we assigned onomatopoeic words to 9,722 audio samples.
In order to collect onomatopoeic words efficiently, we used the crowdsourcing platform Lancers \cite{Lancers}.
Recently, the crowdsourcing platform has often been used to create large-scale datasets \cite{Drossos_ICASSP2020, Lipping_Dcase2019, Takamichi_JVScorpus_2019, Hughes_INTERSPEECH_2010}.

Using the crowdsourcing platform, we asked a crowdworker to conduct the following tasks:
\begin{itemize}
  \setlength{\itemsep}{0pt}
  \item[\bf Task I:] \textbf{Collection of onomatopoeic words and a self-reported confidence score for each audio sample}\\
  After listening to one audio sample, the crowdworker gives three different onomatopoeic words and self-reported confidence scores for each onomatopoeic word.
  The self-reported confidence score is on a scale of five from 1 (very unconfident) to 5 (very confident) for onomatopoeic words.
  Onomatopoeic words were collected from more than five crowdworkers for each audio sample.\\[-6pt]
  \item[\bf Task II:] \textbf{Collection of others-reported acceptance score for onomatopoeic words given to others}\\
  We present an audio sample and onomatopoeic words to the crowdworker.
  The crowdworker gives an others-reported acceptance score for each onomatopoeic word.
   The others-reported acceptance score is on a scale of five from 1 (highly unacceptable) to 5 (highly acceptable) for each onomatopoeic word by others, who were not the worker giving onomatopoeic words.
  The others-reported acceptance score was collected from more than five crowdworkers for onomatopoeic words with 4 or high confidence levels.\\[-10pt]
\end{itemize}

In the case of different onomatopoeic words being given for one audio sample, we can select onomatopoeic words used for environmental sound synthesis using the self-reported confidence score and others-reported acceptance score.
\begin{figure}[t!]
\begin{center}
\includegraphics[width=0.80\columnwidth]{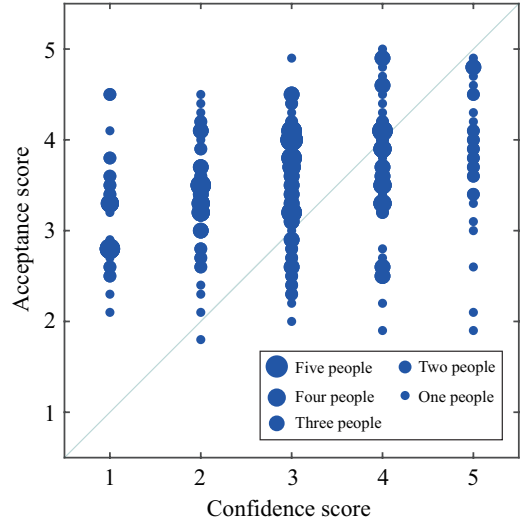}
\vspace{-8pt}
\caption{Scatter plot of self-reported confidence scores and others-reported acceptance scores}
\label{fig:scatter_conf_acceptance}
\vspace{-15pt}
\end{center}
\end{figure}
%
%
\vspace{-3pt}
\section{ANALYSIS OF ONOMATOPOEIC WORDS}
\label{sec:analysis_onomatope}
\vspace{-3pt}
%
%
%
\vspace{-3pt}
\subsection{Collected Onomatopoeic Words}
\label{subsec:collected_onomatope}
\vspace{-3pt}
We discuss the characteristics of the collected onomatopoeic words. 
Table~\ref{table:exapmle_onomatopoeic} shows examples of the collected onomatopoeic words written in {\it katakana} converted to the phoneme representation, self-reported confidence scores, and others-reported acceptance scores.
Since multiple others-reported acceptance scores are given to one onomatopoeic word, Table~\ref{table:exapmle_onomatopoeic} shows the average others-reported acceptance scores.
To express the length of the sound, some workers gave onomatopoeia by repeating the same character, such as ``b u b u b u b u b u''.
\begin{figure}[t!]
\includegraphics[width=1.00\columnwidth]{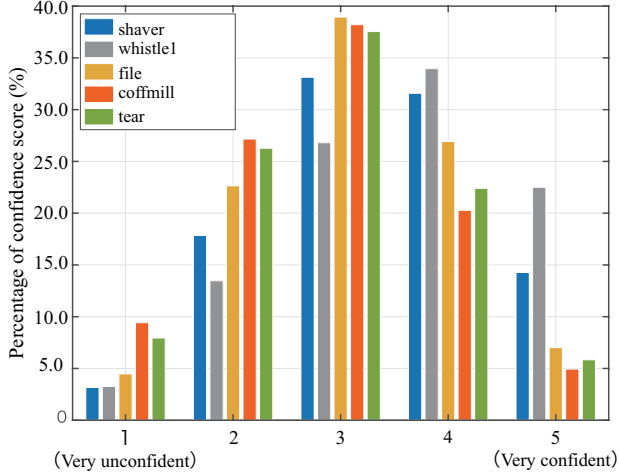}
\vspace{-22pt}
\caption{Percentage of self-reported confidence score for each sound event}
\label{fig:Average_conf_eachevent}
\vspace{-3pt}
\end{figure}
\vspace{-3pt}
\subsection{Collected Self-reported Confidence Scores and Others-reported Acceptance Scores}
\label{subsec:collected_conf_acc}
\vspace{-3pt}
We study the appropriateness of collected onomatopoeic words on the basis of self-reported confidence scores and others-reported acceptance scores. 
To analyze onomatopoeic words using the self-reported confidence and others-reported acceptance scores, we collected the acceptance scores for all onomatopoeic words of 15 types of sound event.
Fig.~\ref{fig:conf_acceptance} shows average self-reported confidence scores and others-reported acceptance scores, and the variance for 15 types of sound event.
The others-reported acceptance score tends to be higher than the self-reported confidence score in Fig.~\ref{fig:conf_acceptance}.
In Fig.~\ref{fig:conf_acceptance}, overall, self-reported confidence and others-reported acceptance scores have high values, and we were able to collect appropriate onomatopoeic words to express audio samples. 

Fig.~\ref{fig:scatter_conf_acceptance} shows the scatter plot of self-reported confidence scores and the average of others-reported acceptance scores.
In Fig.~\ref{fig:scatter_conf_acceptance}, the size of the blue circle indicates the number of samples.
From these results, it seems that the self-reported confidence score tend to be higher than the others-reported acceptance score. 
Therefore, an onomatopoeic word given a high self-reported confidence score is accepted relatively easily by others. 
In addition, despite having low self-reported confidence scores, there are some onomatopoeic words with high others-reported acceptance scores, such as the ``j i:'' sound of an operating electric shaver in Table~\ref{table:exapmle_onomatopoeic}.
There are onomatopoeic words having high self-reported confidence scores despite having low others-reported acceptance scores, such as the ``m i: q'' sound of rubbing a metal rod with a metal file.
Therefore, the use of both the self-reported confidence score and the others-reported acceptance score is very useful for selecting appropriate onomatopoeic words for audio samples.

Figs.~\ref{fig:Average_conf_eachevent} and \ref{fig:Average_acce_eachevent} show the percentage of each score relative to all self-reported confidence scores and all others-reported acceptance scores in each sound event. 
The whistle sound had the highest self-reported confidence score among all sound events in Fig~\ref{fig:Average_conf_eachevent}.
These results show that the whistle sound has high intelligibility \cite{okamoto_arXiv_2019}, and collected onomatopoeic words are similar.
Fig.~\ref{fig:Average_acce_eachevent} indicates that the others-reported acceptance score is relatively higher than the self-reported confidence score.
Additionally, the sound of a whistle had the highest others-reported acceptance score among all sound events, the same as the self-reported confidence score in Fig.~\ref{fig:Average_conf_eachevent}.
Moreover, as seen in Figs.~\ref{fig:Average_conf_eachevent} and \ref{fig:Average_acce_eachevent}, each sound event has a high percentage of confidence and acceptance levels of 3 or higher. 
Thus, we were able to collect many onomatopoeic words to express audio samples.

There are onomatopoeic words with very low self-confidence and others-reported acceptance scores, for example, the ``g i ry a g i ry a" sound by an operating electric shaver in Table~\ref{table:exapmle_onomatopoeic}.
We believe that these onomatopoeic words should be left out of environmental sound synthesis using onomatopoeic words.
These results showed that self-reported confidence scores and others-reported acceptance scores of onomatopoeic words were very useful for selecting onomatopoeic words to express an audio sample used for environmental sound synthesis.

Our dataset include many onomatopoeic words.
We consider that this dataset can be used in a wide variety of research, not just environmental sound synthesis, for example, the generation of onomatopoeic words from audio signals \cite{Ikawa_ICASSP2018_01}.
\begin{figure}[t!]
\includegraphics[width=1.00\columnwidth]{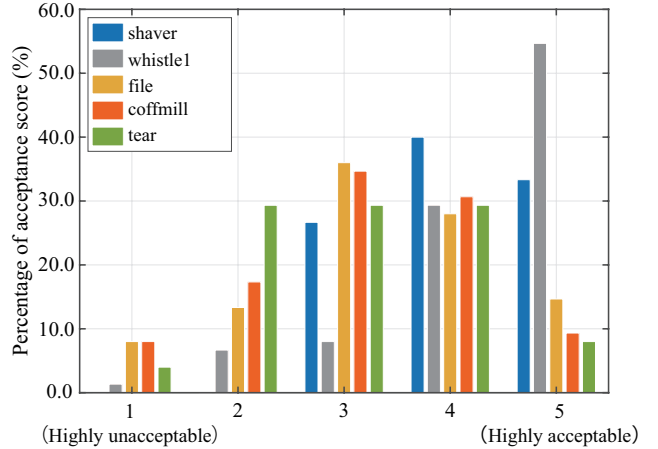}
\vspace{-36pt}
\caption{Percentage of others-reported acceptance score for each sound event}
\label{fig:Average_acce_eachevent}
\vspace{5pt}
\end{figure}
%
\vspace{-3pt}
\section{CONCLUSION}
\label{sec:conclusion}
\vspace{-3pt}
We constructed a dataset for environmental sound synthesis, named RWCP-SSD-Onomatopoeia, containing 155,568 onomatopoeic words, using the crowdsourcing platform.
We also collected each crowdworker's self-reported confidence scores and others-reported acceptance scores for each onomatopoeic word in order to select onomatopoeic words. 
On the basis of the results of collected self-reported confidence scores and others-reported acceptance scores, we were able to collect onomatopoeic words that are acceptable to many people.
We also showed that self-reported confidence scores and others-reported acceptance scores enable us to select onomatopoeic words to express audio samples used for environmental sound synthesis.

In the future, we will conduct environmental sound synthesis using onomatopoeic words included this dataset. 

%
%
%
\vspace{-3pt}
\section{ACKNOWLEDGMENT}
\label{sec:ack}
\vspace{-3pt}
This  work  was  supported  by  JSPS  KAKENHI  Grant  Number JP19K20304 and ROIS NII Open Collaborative Research 2020 Grant Number 20S0401.
%
%
%
\bibliographystyle{IEEEtran}
\bibliography{refs}

\end{sloppy}
\end{document}